# A Taxonomy of Metacognitive Learning Scenarios in Professional Contexts: Integrating Systems Theory with Empirical Constraints

David C Gibson
Curtin University, UNESCO Co-Chair
Perth, Western Australia
Australia
davidcgibson50@gmail.com

Mary Elizabeth Azukas
Georgia Institute of Technology
Atlanta, Georgia
United States
lizazukas@gmail.com

Meryem Yilmaz Soylu
Georgia Institute of Technology
Atlanta, Georgia
United States
meryem@gatech.edu

Correspondence: davidcgibson50@gmail.com

## Abstract

Metacognitive theories provide foundational frameworks for understanding self-regulated learning yet lack systematic integration into comprehensive scenario taxonomies capable of guiding AI-enhanced professional development interventions. Existing models inadequately specify how metacognitive components combine into distinct learning scenarios or how professionals progress from novice to expert functioning. A six-node open systems model (Environment, Input, Processes, Structures, Output, Feedback) was developed by synthesizing four major theoretical frameworks. Combinatorial enumeration generated 216 mathematically possible learning scenarios. Four sequential constraint-based filters—psychological plausibility, educational relevance, measurement feasibility, and intervention potential—informed by empirical workplace learning research, reduced this space to 24 priority scenarios. Five focal scenarios were subjected to a formal concept analysis. The 24 priority scenarios were distributed across three developmental tiers: novice (6), developing (10), and expert/adaptive (8). Analysis revealed critical theoretical gaps regarding the dynamic reconfiguration of monitoring-control relationships across expertise levels, the role of feedback topology in metacognitive development, and trade-offs between internal integration and external connectivity. Multiple viable developmental trajectories were identified. The taxonomy enables targeted, scenario-specific professional development interventions and generates testable predictions for advancing metacognition theory beyond primarily descriptive accounts.

## Introduction

Metacognition, defined as 'thinking about thinking', has emerged as a critical factor in professional learning and development [1, 2]. Foundational theories distinguish between metacognitive *knowledge* (what learners know about cognition) and metacognitive *regulation* (how learners control their cognitive processes), with monitoring and control identified as core functional mechanisms [3, 4]. However, despite decades of research establishing metacognition's importance for learning outcomes, the field lacks a systematic taxonomy that integrates these theoretical components into comprehensive scenario frameworks capable of guiding targeted research and interventions in AI-enhanced professional and lifelong learning contexts.

Professional learning environments present unique challenges for metacognitive theory. Unlike formal educational settings where metacognitive research has flourished, workplace learning is characterized by informality, contextual variability, competing demands, and the necessity for autonomous self-regulation [5, 6]. Empirical studies confirm that metacognitive strategies, including planning, monitoring, and regulation, operate in workplace settings and correlate with learning effectiveness [7], yet theoretical models derived primarily from K-12 and higher education contexts may not fully capture the distinctive patterns of professional metacognition.

This paper addresses three interrelated theoretical gaps. First, existing metacognition frameworks describe components and processes in relative isolation, lacking systematic integration that specifies how these elements combine to form distinct learning scenarios situated in realistic problem-solving contexts. Second, the field has not systematically examined which of the many theoretically possible metacognitive patterns occur with sufficient frequency to warrant pedagogical attention in professional contexts. Third, current theories inadequately account for developmental trajectories, the pathways through which professionals progress from novice to expert metacognitive functioning.

We propose a refined taxonomy generated through systems modeling that addresses these gaps by: (1) integrating major metacognitive theories [1, 2, 8, 9] into a unified six-node open systems framework, (2) systematically defining and enumerating possible learning scenarios, (3) applying empirically-informed constraints to identify priority patterns, and (4) conducting deep analysis of focal scenarios to reveal theoretical limitations and generate testable predictions. Our analysis focuses specifically on professional learning contexts while acknowledging implications for lifelong learning trajectories.

## Theoretical Framework: Systems Model and Historical Theories

The proposed system model synthesizes four major theoretical frameworks into a six-node architecture based on open systems theory as implemented in the DOT Framework [10]: **Environment** (external context and task demands), **Input** (information entering the learner's system), **Processes** (active cognitive/metacognitive operations), **Structures** (stable knowledge representations, beliefs, standards), **Output** (generated products and behaviors), and **Feedback** (information about output quality and environmental response). This architecture provides a common language for mapping and analyzing diverse theoretical contributions.

Flavell's [1, 11] framework contributed the fundamental distinction between metacognitive knowledge and metacognitive experiences, mapped onto our model as Structures (stable knowledge about cognition) and the dynamic interplay between Processes and Structures (online metacognitive experiences during task engagement). Flavell's four components (metacognitive knowledge, experiences, goals, and actions) are distributed across our system as follows: knowledge resides in Structures; experiences emerge from Processes-Structures interactions; goals are encoded as standards within Structures; and actions manifest as control operations flowing on the links from Structures to Processes.

Nelson and Narens' [2, 12] two-level model provided the clearest specification of functional mechanisms. Their object-level and meta-level architecture maps directly onto the bidirectional links between Processes (object-level) and Structures (meta-level). Critically, Nelson and Narens specified *monitoring* as upward information flow (Processes → Structures) and *control* as downward information flow (Structures → Processes), establishing the functional basis for metacognitive regulation. The use of the arrow in our computational representation means

“flows to, influences, or causes a change in” the receiving node. Bidirectionality is captured in our model through three possible internal arrangements: P→S (bottom-up monitoring dominates), S→P (top-down control dominates), or P⇄S (fully integrated bidirectional processing). Appendix 1 contains additional details of the computational representation’s symbolic system. In the fully integrated mode, both nodes take turns influencing each other over time, so both *directionality* and *precedence* are involved, which has consequences for causal modeling [13].

**Figure 1**
*Open system model of cognition*

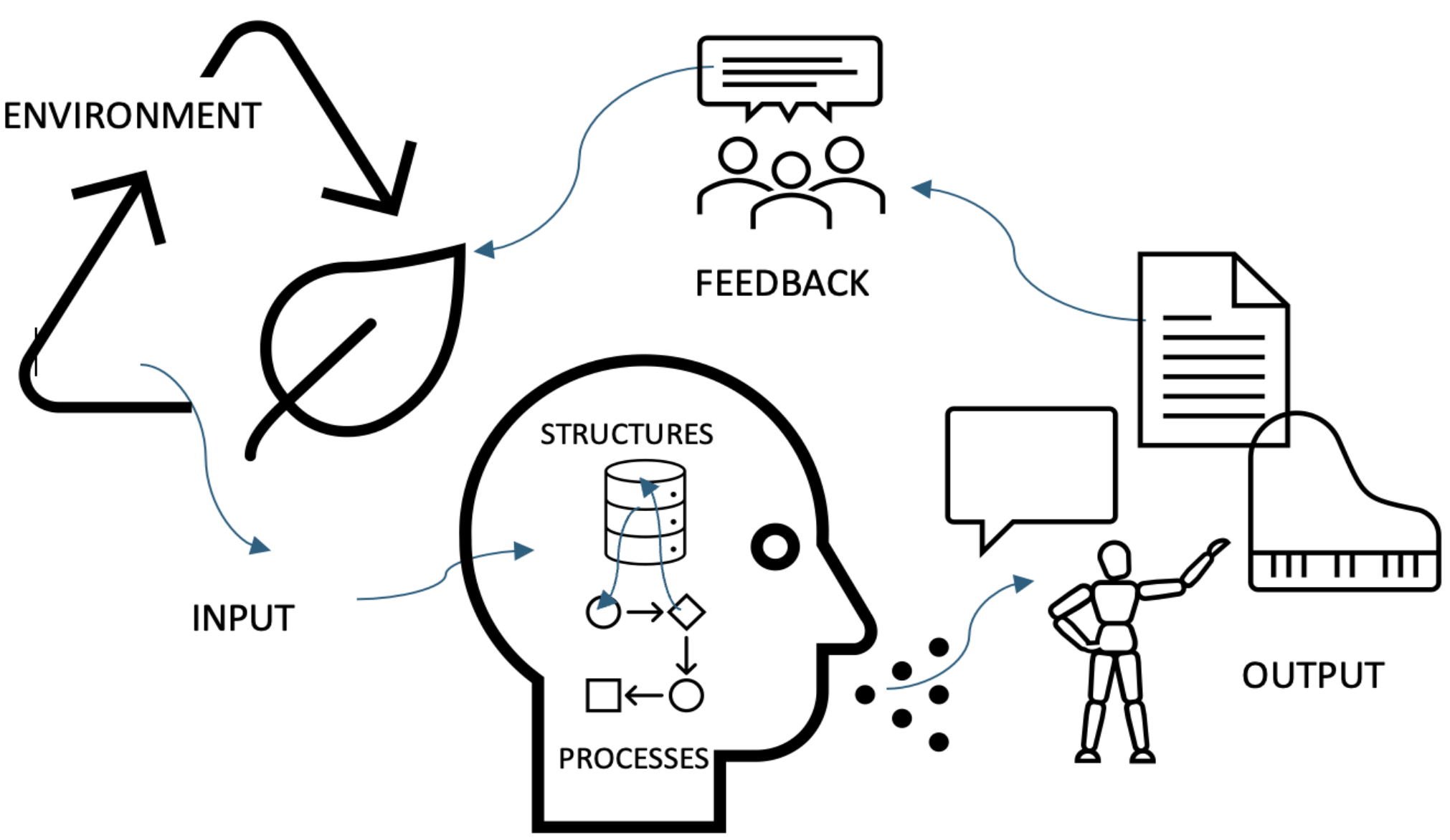

NOTE: The open systems model of metacognition has six nodes. Four external (OUTPUT, FEEDBACK, ENVIRONMENT, INPUT) and two internal (PROCESSES and STRUCTURES), and seven primary links or relationships.

Koriat's [8] cue-utilization framework refined the understanding of monitoring mechanisms by distinguishing experience-based monitoring (drawing on processing fluency and noetic feelings) from information-based monitoring (drawing on retrieved knowledge and beliefs). In our systems model, experience-based monitoring manifests as immediate Processes → Structures links triggered by online processing dynamics, while information-based monitoring involves Structures → Processes links where existing knowledge shapes metacognitive judgments. This distinction becomes particularly important in understanding how professionals develop increasingly sophisticated monitoring capabilities.

The COPES model [9] expanded the theoretical framework of metacognition by adding external structures and processes, like external task standards and instructor feedback, to the historically primary focus on internal mechanisms. Importantly, the COPES model provided a foundation to specify self-regulated learning within a distributed cognitive system. Their five facets: Conditions, Operations, Products, Evaluations, and Standards, map systematically onto the open system nodes: Conditions correspond to Structures (knowledge) as well as Input (task information); Operations correspond to Processes; Products correspond to Output; Internal evaluations emerge from the monitoring flow (Processes → Structures); and Standards reside in Structures as comparison criteria and in Environment as professional standards of practice. The COPES model's emphasis on IF-THEN production rules implementing metacognitive control aligns with our conception of Structures → Processes links as conditional control mechanisms. See [14] for a detailed analysis of COPES from an open systems perspective.

### Cross-Cluster Connections and External Topology

A critical innovation of the systems model is the explicit representation of **cross-cluster connections,** the interface between the internal metacognitive system, symbolized in our notation within brackets and highlighted in gray in the taxonomy (Appendix 1), for example, [(Processes and Structures)] to distinguish it from the external learning environment (Environment, Input, Output, Feedback). We constrain the cross-cluster connections to preserve psychological realism: Input always flows into the internal system (Input → [Processes] and/or Input → [Structures]), while Output always flows from the internal system ([Processes →] Output and/or [Structures →] Output). This constraint reflects the fundamental principle that metacognitive systems mediate between environmental information and behavioral responses.

The **external topology,** the arrangement of Environment, Output, Feedback, and Input nodes, represents the structure of the external part of the learning loop. Empirical evidence from workplace learning suggests that feedback processes are critical for professional development [15], yet feedback mechanisms vary considerably across contexts. We identified a mandatory backbone sequence: Output → Feedback → Environment → Input (O→F→E→I), representing the realistic flow wherein professional output into a professional community must receive feedback, and be processed through the environmental context of the community of practice, before becoming new input for subsequent learning cycles [16, 17]. However, this backbone admits multiple "shortcuts" where direct connections bypass intermediate stages: for example, Output → Input (self-monitoring without environmental mediation), Feedback → Input (direct feedback incorporation), Output → Environment (immediate environmental impact), creating eight distinct external topologies for computational modeling (see Appendix 2).

Systems integration reveals a previously unrecognized theoretical gap: existing frameworks inadequately specify how internal metacognitive architecture (the arrangement of Processes and Structures) interacts with a variety of external feedback topologies (the arrangement of environmental learning loops). For instance, Nelson and Narens' model [2] elegantly describes monitoring and control within the internal system but remains largely silent on how external feedback integrates with internal monitoring processes. Similarly, [9] addresses external conditions but provides limited theoretical guidance on how different feedback structures might differentially support or hinder metacognitive development. Our systematic enumeration forces explicit consideration of these interactions.

## Methodology: Scenario Generation and Filtering

### Combinatorial Generation

The six-node systems model with specified constraints generates 216 distinct scenarios through combinatorial mathematics (a simple directed graph of six nodes has 6! or 720 unique pathways). We used Claude Sonnet 4.5 in extended thinking and research modes to formalize the data structures, conduct formal concept analysis of the network, and link the metacognitive literature to the model. The model concept and frameworks come from [10], which was extended with additional theories and processes during work on an AI-enabled Coach focused on helping users understand and use the core frameworks.

Internal cluster arrangements. The Processes and Structures nodes can be arranged in three configurations: (1) P→S (bottom-up: processes inform structures through monitoring), (2) S→P (top-down: structures control processes through regulation), or (3) P⇄S (bidirectional: integrated monitoring and control). This yields three internal arrangements.

Cross-cluster connection patterns**.** Input can connect to Processes, Structures, or both, yielding 3 entry patterns. Similarly, Output can originate from Processes, Structures, or both, yielding 3 exit patterns. The combination produces nine cross-cluster patterns (3 × 3).

External topologies**.** The mandatory O→F→E→I backbone admits a range from zero to three additional shortcuts: (1) O→E (output directly influences environment), (2) O→I (immediate self-monitoring), (3) F→I (direct feedback incorporation). Including the baseline with no shortcuts, this generates eight external topologies ($2^3$ possibilities).

The total scenario space equals 3 × 9 × 8 = 216 mathematically valid scenarios, each representing a distinct configuration of how internal metacognitive processes, cross-system connections, and external feedback structures combine to constitute a learning episode. Appendix 1 provides a complete glossary of terms and relationships, and Appendix 2 provides a selection of 24 scenarios created via a four-stage filtering of constraints, as discussed next.

**Four-Stage Constraint-Based Filtering**

We applied four sequential filters to reduce the 216-scenario space to priority patterns warranting additional theoretical and empirical attention to enable computational modeling of professional learning contexts so the AI Coach would generate sound advice. Each filter was informed by empirical evidence from workplace learning research and grounded in explicit psychological and pedagogical reasoning.

**Filter 1: Psychological Plausibility (216 reduced to 178 scenarios).** This filter eliminated scenarios containing logical inconsistencies or psychologically implausible sequences. For example, configurations where Input → (flows into) Structures, and Processes → (flow into) Output but with internal arrangement where Processes only flow to Structures (P→S) were eliminated because they create a disconnected flow: if input activates only knowledge, and processes inform structures bottom-up, yet output derives only from processes never informed by the activated knowledge, the scenario violates basic information-processing principles. Empirical research confirms that metacognitive strategies and cognitive processing operate in concert rather than in isolated sequences [7]. Similarly, we eliminated pure knowledge-pathway scenarios (Input → Structures, Structures → Output, with internal S→P) that bypass all process involvement, as professional learning research demonstrates that even declarative knowledge responses require some degree of cognitive processing [18]. This filter removed 38 scenarios containing structural impossibilities or patterns unsupported by cognitive theory.

**Filter 2: Educational Relevance (178 reduced to 141 scenarios).** This filter prioritized scenarios amenable to workplace interventions and professional development initiatives. We eliminated patterns that, while psychologically possible, represent either missed developmental opportunities or organizationally impractical configurations. For instance, scenarios combining fully integrated internal processing (P⇄S) with minimal external feedback utilization (baseline O→F→E→I topology only) were eliminated because empirical evidence demonstrates that professionals with integrated metacognitive systems actively utilize multiple feedback channels [15]. Retaining such scenarios would be pedagogically counterproductive as they represent suboptimal rather than target patterns. Additionally, we eliminated patterns requiring organizational restructuring that empirical research suggests would be rejected by employees or prove impractical to implement [19]. This filter removed 37 scenarios lacking educational value despite psychological plausibility.

**Filter 3: Measurement Feasibility (141 reduced to 80 scenarios).** This filter addressed a critical methodological constraint: scenarios must be empirically distinguishable using available assessment instruments. Current workplace learning measures, including the Self-Regulated Learning at Work Questionnaire [20], think-aloud protocols, and learning analytics trace data, cannot reliably distinguish internally isolated fine-grained temporal sequences or micro-level processing differences. For example, whether Input activates Processes microseconds *before* Structures versus *simultaneously* cannot be detected with existing methods, without specialized laboratory equipment in an inauthentic experimental setting. We therefore consolidated or eliminated scenarios differing only in undetectable micro-sequences. This pragmatic filter acknowledges that theoretical distinctions lacking operational definitions remain scientifically untestable for most practical purposes like organizational change management, workplace or cognitive coaching in realistic settings. Importantly, this consolidation reveals another theoretical gap: current metacognition theories postulate finer-grained distinctions than existing measurement technologies can validate in authentic performance contexts. This filter removed 61 scenarios that, while theoretically distinct, would appear empirically identical in workplace research studies.

**Filter 4: Intervention Potential (80 reduced to 24 scenarios).** The final filter selected scenarios representing clear developmental targets where training, coaching, or instructional design could facilitate meaningful change. Informed by meta-analytic evidence showing that metacognitive strategy instruction produces medium to large effects on learning [21], we prioritized scenarios along three dimensions: (1) **developmental progression:** scenarios representing teachable transition points from novice to expert functioning, (2) **intervention responsiveness:**patterns where empirical evidence demonstrates training effectiveness, and (3) **workplace relevance:** configurations aligning with organizational needs for innovation, adaptation, and continuous learning [22]. This criterion-based selection yielded 24 priority scenarios organized into three developmental tiers: novice professional learners (6 scenarios), developing professionals (10 scenarios), and expert/adaptive professionals (8 scenarios). A complete list of the 24 scenarios is provided in Appendix 2.

### Empirical Grounding

Each filtering decision was anchored in empirical evidence from professional and workplace learning research. For psychological plausibility, we drew on studies demonstrating which information-processing sequences occur in adult learners [7]. For educational relevance, we consulted meta-analyses that identified which metacognitive interventions prove effective in professional contexts [20]. For measurement feasibility, we examined the capabilities and limitations of validated workplace learning instruments. For intervention potential, we synthesized evidence on which patterns respond to training and which organizational factors support metacognitive development [6]. This grounding distinguishes the taxonomy from purely logical enumerations, ensuring that priority scenarios reflect empirical realities of professional learning.

## Results: Priority Scenarios and Their Characteristics

The 24 priority scenarios are distributed across three developmental tiers, each characterized by distinct metacognitive profiles and external engagement patterns. Appendix 2 presents the complete taxonomy with abbreviated notation, descriptions, and example references.

### Tier 1: Novice Professional Learners (Scenarios 1-6)

Novice scenarios are characterized by unidirectional internal processing (predominantly P→S or S→P), limited cross-cluster connectivity (single entry or single exit points), and partial external engagement (baseline external topology plus one or two shortcuts). These patterns reflect learners who are developing basic metacognitive monitoring capabilities but have not yet achieved fully integrated self-regulation.

Scenario 1 (I→[P, P→S, P→]O, O→F→E→I + F→I) for example, represents the foundational learning pattern: experience-driven processing that updates metacognitive knowledge through bottom-up monitoring, with direct feedback incorporation facilitating learning. This aligns with experience-based monitoring [8], where learners rely on processing fluency and noetic feelings to generate metacognitive judgments. The presence of the F→I shortcut indicates developing awareness of performance feedback, though the learner has not yet developed robust self-monitoring capabilities (no O→I shortcut).

Scenario 6 (I→[P, P⇄S, P→]O, O→F→E→I + O→I) marks the emergence of bidirectional integration, representing a critical developmental transition. The addition of self-monitoring (O→I) alongside maintained bottom-up processing (input through Processes) indicates nascent metacognitive sophistication. However, the restriction to single output source (Processes only) suggests the learner has not yet learned to flexibly deploy both procedural execution and declarative knowledge in task performance.

**Tier 2: Developing Professionals (Scenarios 7-16)**

Developing professional scenarios exhibit bidirectional or flexible internal processing, increased cross-cluster complexity (often dual entry or dual exit), and enhanced external engagement (typically 2-3 shortcuts active). These patterns characterize learners who have achieved basic metacognitive competence and are refining their self-regulatory capabilities.

Scenario 7 (I→[P, P⇄S, {P,S}→]O, O→F→E→I + O→I) emerges as the empirically most common and theoretically most significant pattern for developing professionals. The fully integrated internal system (P⇄S) enables dynamic interplay between online processing and stored knowledge, consistent with [2] bidirectional monitoring-control architecture. The dual-source output ({P,S}→O) indicates flexible performance drawing on both procedural skills and declarative knowledge. The presence of self-monitoring (O→I) alongside maintained environmental feedback suggests balanced reliance on internal and external evaluation sources. Empirical evidence from workplace learning strongly supports this pattern: studies show that metacognitive monitoring serves as the "gateway to self-regulation" [23], and professionals demonstrating such integrated functioning show superior learning outcomes [20].

Scenario 14 (I→[{P,S}, P⇄S, {P,S}→]O, O→F→E→I + O→I) represents full internal integration: bidirectional processing with parallel entry and exit across both Processes and Structures. This pattern characterizes the upper boundary of Tier 2, indicating sophisticated but not yet maximally optimized metacognitive functioning. The restriction to baseline external topology plus O→I only (lacking F→I and O→E shortcuts) suggests these learners effectively self-monitor but have not yet fully leveraged all available feedback channels.

**Tier 3: Expert/Adaptive Professionals (Scenarios 17-24)**

Expert scenarios are distinguished by consistently bidirectional internal processing (P⇄S), maximal cross-cluster engagement (dual or parallel entry and exit), and rich external connectivity (typically 3-4 shortcuts active, sometimes including the fully connected topology 8 in Appendix1. These patterns reflect highly developed metacognitive sophistication characteristic of expert professional learners.

Scenario 17 (I→[{P,S}, P⇄S, {P,S}→]O, O→F→E→I + O→I + F→I) represents the prototypical expert pattern. All components of the internal system are fully engaged and bidirectionally connected. Both self-monitoring (O→I) and direct feedback incorporation (F→I) are active, enabling rapid calibration of metacognitive judgments against both internal standards and external criteria. This pattern aligns with research on expert performance showing that top performers systematically self-observe while simultaneously processing external performance information [24].

Scenario 18 (I→[{P,S}, P⇄S, {P,S}→]O, Full topology 8 in Appendix 1 represents maximum integration: the fully connected external topology means all possible feedback pathways are active simultaneously. While empirically rare due to cognitive load constraints, this pattern may characterize peak performance states or highly automatized expertise where parallel processing of multiple feedback streams becomes possible.

**Cross-Cutting Patterns and Theoretical Insights**

Analysis of the 24 priority scenarios reveals several patterns not explicitly predicted by existing theories:

*Feedback topology matters differentially across development.* Novice scenarios tolerate simpler external topologies (1-2 shortcuts), while expert scenarios consistently employ richer connectivity (2-4 shortcuts). This suggests that metacognitive development involves not just internal integration (P⇄S) but also increasingly sophisticated environmental engagement, a dimension inadequately theorized in current frameworks focused primarily on internal monitoring-control processes.

*Internal arrangement and cross-cluster patterns covary systematically.* Scenarios with S→P internal arrangements (top-down control) more frequently feature I→S entry patterns (knowledge activation first), while P→S scenarios more frequently feature I→P entry patterns (processing first). This covariation suggests initial information routing may prime subsequent internal dynamics, a possibility not addressed in existing theories that treat entry patterns and internal processing as independent.

*The "missing middle" problem*. The filtering process eliminated numerous scenarios occupying intermediate positions between clear developmental stages. This suggests that metacognitive development may involve discrete shifts rather than continuous gradual progression, consistent with dynamic systems theories but at odds with linear developmental models implicit in much SRL literature.

A review of the literature suggests a set of five scenarios that have received the most attention in research and practice.

**Table 1**

*Scenarios commonly found in the metacognitive literature*

| Scenario | Rationale |
|---|---|
| I→ [P, P→S, P→] O, O→F→E→I + F→I | Foundational Experience-Based Learning; Bottom-up learning with direct feedback |
| I→ [P, P⇄S, {P,S}→] O, O→F→E→I | Standard learning cycle with full integration |
| I→ [S, S→P, P→] O, O→F→E→I + O→I | Application of knowledge with self-monitoring |
| I→ [{P,S}, P⇄S, {P,S}→] O, O→F→E→I + O→I | Expert performance with full awareness |
| I→ [S, S→P, S→] O, O→F→E→I | Pure knowledge application (routine tasks) |

These well-defined metacognitive pathways have been the focus of research for perhaps three reasons. Most of the models emphasize metacognitive monitoring as a gateway to self-regulation [25], suggesting that P⇄S (i.e. multiple connections between processes and structures) is common. Second, learners tend to rely on both domain and metacognitive knowledge [26], suggesting I→{P,S} (i.e. inputs from the environment go directly and simultaneously to both processes such as senses, or to structures, such as assimilation matching with memory). Third, self-monitoring (the O→I shortcut, where the learner makes immediate reflective use of their output without feedback from others ) is developmentally important and a teachable self-regulation skill [27].

## Analysis: Five Focal Scenarios and Theoretical Gaps

We now conduct in-depth analysis of five scenarios spanning the developmental continuum: Scenario 1 (novice), Scenario 7 (developing standard), Scenario 13 (developing specialized), Scenario 19 (expert bottom-up), and Scenario 24 (integrated expert).

### Scenario 1: Foundational Experience-Based Learning

Configuration: I→[P, P→S, P→]O, O→F→E→I + F→I

This scenario represents the entry point for metacognitive development in professional contexts. Information enters through active processing (I→P), meaning the learner engages with tasks experientially rather than through prior knowledge activation. Bottom-up monitoring (P→S) captures the essence of experience-based metacognition [8]: processing dynamics generate feelings of fluency, disfluency, or knowing that inform metacognitive judgments stored in Structures. The single-process output (P→O) indicates performance driven primarily by execution rather than strategic knowledge application.

*Theoretical gaps revealed*. Existing theories inadequately specify how experience-based monitoring develops from implicit feelings to explicit metacognitive knowledge. The [8] framework describes the phenomenology of experience-based cues but provides a limited developmental account. Our scenario suggests this transition requires repeated cycles through the entire system model, where P→S monitoring patterns crystallize into stable Structures that can subsequently exert S→P control. The repetition via complete cycling is supported by both neurological evidence of Hebbian training ("connections that fire together, wire together") and the cognitive science principle of "spaced repetition" [28]. However, none of the major metacognitive theories provide detailed mechanisms for this crystallization process, such as the detailed physical model of learning in [14].

The F→I shortcut indicates direct feedback incorporation, yet the historical metacognition theories don't distinguish between passive feedback reception versus active feedback-seeking. Professional learning research shows that help-seeking and feedback-seeking are distinct metacognitive strategies [7], but our systems model treats F→I as a single connection type. This suggests the need in our model for finer-grained specification of feedback processes, potentially distinguishing *pushed* feedback (environment-initiated) from *pulled* feedback (learner-initiated), with implications for practice in workplaces.

Developmental predictions. Learners in Scenario 1 should:

- Show high reliance on trial-and-error learning
- Demonstrate limited planning (minimal S→P control)
- Benefit from structured external feedback (strong F→I link)
- Struggle with transfer to novel tasks (limited Structures to generalize from)

**Scenario 7: Core Integrated Learning**

**Configuration:** I→[P, P⇄S, {P,S}→]O, O→F→E→I + O→I

This scenario emerged from filtering as the predicted most common pattern for developing professionals. The bidirectional P⇄S arrangement enables the full monitoring-control cycle as theorized [2]: processes inform the meta-level through monitoring, which then guides object-level processing through control. The dual-source output ({P,S}→O) represents a critical advance: the learner can now flexibly deploy both procedural execution and strategic knowledge application depending on task demands.

The presence of self-monitoring (O→I) alongside maintained experiential entry (I→P) creates what we term the "reflect-while-doing" configuration: the learner processes tasks experientially but simultaneously monitors their own output without waiting for environmental feedback. This pattern aligns with a conception of monitoring as occurring within phases of self-regulated learning, not just between phases [9].

*Theoretical gaps revealed*. No existing theory adequately explains how learners coordinate multiple monitoring sources (both internal P→S monitoring of processing dynamics AND external O→I monitoring of output quality). Do these operate independently, sequentially, or in integrated fashion? One framework [8] addresses monitoring source (experience vs. information) but not monitoring multiplicity. [9] describes monitoring-based standards comparisons but doesn't specify how multiple comparison processes interact.

The dual-source output ({P,S}→O) raises a second theoretical question: how do learners decide whether to generate output from Processes or Structures? Existing theories treat this as a domain-general control decision, but our analysis suggests it may be scenario-specific. In Scenario 7, both sources are active, implying either parallel output generation or rapid switching mechanisms—neither of which current theories specify in detail.

Developmental predictions. Learners in Scenario 7 should:

- Demonstrate balanced monitoring (both internal and external)
- Flexibly adapt performance strategies (dual-output capability)
- Show emerging but incomplete self-regulation (still rely on external feedback)
- Benefit from metacognitive prompting that highlights P⇄S connections

*Lifelong learning implications*. Scenario 7 may represent a stable attractor state for many professionals—sufficient metacognitive sophistication for effective routine performance, but not the enhanced capabilities required for expert adaptation. This suggests that many professionals may plateau at Scenario 7 unless specific interventions facilitate transition to richer external engagement patterns (Scenarios 13-17).

**Scenario 13: Specialized Developing Pattern**

**Configuration:** I→[{P,S}, S→P, P→]O, O→F→E→I + O→I + F→I

This scenario presents a developmental variant: top-down control (S→P) combined with parallel entry (I→{P,S}) and enhanced feedback connectivity (both O→I and F→I active). Unlike Scenario 7's experiential emphasis, Scenario 13 reflects a more knowledge-driven approach where existing schemas and beliefs guide processing.

The parallel entry pattern (I→{P,S}) indicates simultaneous activation of both processing and knowledge systems upon encountering new information. This aligns with dual-process theories but raises an important question: if both Processes and Structures receive input simultaneously, yet Structures control Processes (S→P), what prevents circular or contradictory signals? Existing theories don't address this potential conflict.

The combination of O→I and F→I shortcuts creates what we term "dual-channel calibration": the learner compares self-monitored output against internal standards (O→I) while also integrating external feedback (F→I). This pattern is particularly relevant for professional contexts requiring rapid self-correction (self-monitoring) combined with external validation (feedback integration).

*Theoretical gaps revealed.* Current theories inadequately address the relationship between control direction (top-down S→P) and entry pattern (parallel I→{P,S}). If knowledge structures exert control but information enters both systems simultaneously, the theories need to specify precedence rules or conflict resolution mechanisms. One possibility here is that 'fast thinking' has precedence over 'slow thinking' unless the context is low stress and unrushed [29, 30].

Additionally, Scenario 13 reveals the "single-source output paradox": despite parallel entry and dual feedback channels, output derives only from Processes (P→O). This seems inefficient—why activate Structures if they don't contribute to output? One possibility is that Structures serve a purely regulatory role in this scenario, generating metacognitive evaluations but not content. However, existing theories don't clearly distinguish between Structures as content-generators (active control?) versus Structures as regulators (passive control?).

Developmental predictions**.** Learners in Scenario 13 should:

- Demonstrate strong planning and strategic thinking (S→P control)
- Show rapid self-correction (dual feedback channels)
- Potentially exhibit over-reliance on existing knowledge (top-down bias)
- Struggle when existing knowledge is inadequate (limited P→S updating)

*Lifelong learning implications***.** Scenario 13 may characterize mid-career professionals with well-developed domain knowledge who rely heavily on experience-based schemas. The limitation is that top-down control without bidirectional integration (no P⇄S) may impede knowledge updating, potentially leading to expertise rigidity. Interventions should focus on enhancing bottom-up monitoring (adding P→S alongside S→P) to enable schema revision.

**Scenario 19: Expert Bottom-Up Pattern**

**Configuration:** I→[{P,S}, P→S, {P,S}→]O, O→F→E→I + O→I + F→I + O→E

This scenario represents an expert pattern: despite advanced external connectivity (4 shortcuts including the rare O→E direct environmental influence), the internal arrangement remains unidirectional bottom-up (P→S). This contradicts the common assumption that expertise requires bidirectional P⇄S integration.

The parallel entry (I→{P,S}) and dual-source output ({P,S}→O) indicate full engagement of both processing and knowledge systems. However, the P→S arrangement suggests that in this pattern, metacognitive knowledge serves primarily as a monitoring repository rather than a control mechanism. This aligns with certain theories of implicit learning and skilled performance where control becomes automated and metacognitive awareness serves retrospective sense-making rather than online regulation [31].

The inclusion of O→E (direct environmental influence) is particularly notable. This shortcut means the learner's output directly shapes the environment before feedback is generated. This pattern may characterize creative or innovative professionals whose work actively restructures their problem space, rather than responding to pre-existing task demands. On-demand performance of a musician, for example, may be an example.

*Theoretical gaps revealed***.** Existing metacognition theories implicitly privilege P⇄S bidirectionality as the hallmark of expertise, yet Scenario 19 suggests that some forms of expert performance may rely on unidirectional monitoring with highly sophisticated external engagement. This challenges the universality of current metacognitive development models.

The four-shortcut external topology (most connected, excluding the full topology 8 in Appendix 1) combined with unidirectional internal processing suggests compensation mechanisms: rich environmental connectivity may substitute for internal bidirectionality in certain contexts. Current theories don't address such trade-offs between internal integration and external engagement.

Developmental predictions**.** Learners in Scenario 19 should:

- Demonstrate expert performance in specific domains
- Show strong monitoring and accurate calibration (rich feedback use)
- Potentially lack flexible strategy switching (unidirectional S→P absent)
- Excel in creative or generative tasks (O→E environmental restructuring)

*Lifelong learning implications.* Scenario 19 may represent a "specialist expert" trajectory where professionals develop deep proficiency in particular domains through extensive experience-based learning (P→S monitoring) combined with sophisticated environmental engagement, but without developing the strategic flexibility associated with bidirectional control. Returning to the example of a musician, the on-stage performance from memory of a complex composition, requiring highly structured, exact reproduction, with a degree of emotional interpretability, might be an example of this scenario. This suggests multiple pathways to expertise, not all requiring the same internal metacognitive architecture—a possibility not addressed in current theories that assume convergence on P⇄S integration.

**Scenario 24: Integrated Expert Performance**

**Configuration:** I→ [{P,S}, P⇄S, S→] O, O→F→E→I + O→I + F→I + O→E

This scenario represents one form of maximally developed metacognitive functioning. The bidirectional P⇄S integration enables full monitoring-control cycling. Parallel entry (I→{P,S}) ensures comprehensive information processing. Four-shortcut external topology provides maximum feedback connectivity. However, the scenario exhibits a distinctive feature: single-source output from Structures only (S→O).

This pattern suggests knowledge-driven output generation: despite active processing (necessary for P→S monitoring), the actual products generated derive from structured knowledge rather than online processing. This may characterize conceptual expertise where professionals synthesize information into high-level principles, models, or frameworks (Structures) that constitute their primary outputs, rather than producing process-based procedural executions.

The S→O pattern combined with O→E creates a distinctive flow: knowledge generates output that directly reshapes the environment, with multiple feedback channels informing both immediate self-monitoring (O→I) and subsequent knowledge refinement (via P→S after feedback is processed). This pattern may characterize thought leaders, consultants, or strategic planners whose primary professional output consists of conceptual frameworks and recommendations.

*Theoretical gaps revealed.* Existing theories inadequately distinguish between process-driven versus knowledge-driven output generation. One framework [2] assumes outputs are generated at the object level (Processes), yet Scenario 24 demonstrates that outputs can be derived from the meta-level (Structures). This requires theoretical extension: how do meta-level representations become executable outputs without object-level mediation?

Additionally, Scenario 24 highlights an under-theorized aspect of professional metacognition: the role of metacognitive knowledge as generative versus regulatory. Most theories emphasize the regulatory function (monitoring and control), but Scenario 24 suggests that in expert professional contexts, metacognitive knowledge may serve primarily generative functions—creating new conceptual frameworks rather than merely regulating cognitive processes.

Developmental predictions**.** Learners in Scenario 24 should:

- Generate primarily conceptual or strategic outputs
- Demonstrate exceptional metacognitive awareness
- Show strong influence on their professional environments
- Potentially struggle with rapid tactical execution (no P→O direct output)

*Lifelong learning implications.* Scenario 24 may represent an "adaptive expert" endpoint for knowledge-intensive professions requiring continuous conceptual innovation. The developmental pathway likely involves extended periods in P→O or {P,S}→O scenarios before transitioning to pure S→O output generation. This transition marks the shift from doing the work to conceptualizing how the work should be done, a meta-level career transition inadequately addressed in current learning theories focused primarily on task-level performance.

## Discussion: Implications for Lifelong Professional Learning

The five focal scenarios analyzed above reveal systematic patterns in how metacognitive sophistication develops across professional careers. We identify three key developmental dimensions: **(1) internal integration** (unidirectional → bidirectional processing), **(2) cross-cluster complexity** (single → parallel entry/exit), and **(3) external connectivity** (minimal → maximal feedback utilization). Critically, these dimensions do not develop uniformly; different professional trajectories may emphasize different dimensions.

### Developmental Trajectories

The most common trajectory appears to follow: **Scenario 1 → Scenario 7 → Scenario 17/24**, representing progression from experience-based novice learning through core integrated development to expert adaptive functioning. However, our analysis reveals alternative trajectories:

*Specialist pathway.* Scenario 1 → Scenario 6 → Scenario 19, emphasizing bottom-up monitoring and rich external engagement without full bidirectional integration. This may characterize technical specialists whose expertise develops through extensive practice and feedback without requiring strategic flexibility.

*Strategic pathway.* Scenario 3 → Scenario 13 → Scenario 24, emphasizing top-down control and knowledge-driven performance. This may characterize consultants, executives, or strategic planners whose work centers on conceptual innovation and frameworks.

The existence of multiple viable trajectories challenges linear developmental models and suggests that metacognitive development may exhibit equifinality**,** multiple pathways leading to effective professional performance. Current theories inadequately account for such diversity.

### Critical Transition Points

Analysis of the 24-scenario taxonomy identifies several critical transition points where learners may become stuck.

The bidirectionality barrier (Scenarios 1-6 → 7-16). Developing P⇄S integration requires learners to add the missing direction to their existing processing (either adding control to monitoring-dominant patterns or adding monitoring to control-dominant patterns). Empirical evidence suggests this transition is challenging and may require explicit instruction [21].

The self-monitoring threshold (adding O→I shortcut). Scenarios lacking self-monitoring rely entirely on external feedback for performance evaluation. Developing O→I capability—the ability to accurately assess one's own output—represents a critical metacognitive milestone enabling autonomous self-regulation.

The external engagement ceiling (Topology 1 → 2-4 shortcuts). Many professionals may function effectively with minimal external topology (1-2 shortcuts) and never develop richer feedback utilization patterns. This represents a plateau where further development requires active intervention to demonstrate the value of additional feedback channels.

### Implications for Professional Development Interventions

The refined taxonomy enables targeted intervention design based on learners' current scenarios. For learners in **Scenario 1** (novice): Interventions should provide rich, structured feedback (strengthen F→I) and scaffold self-monitoring development (introduce O→I) before attempting to develop bidirectional integration. For learners in **Scenario 7** (developing standard): Interventions should focus on diversifying entry patterns (adding I→S for knowledge activation alongside I→P experiential processing) and enhancing external connectivity (adding F→I and O→E shortcuts). For learners in **Scenario 13** (developing specialized): Interventions should emphasize bottom-up monitoring to complement existing top-down control, preventing expertise rigidity through enhanced P→S knowledge updating. For learners in **Scenario 19** (specialist expert): Interventions might focus on developing bidirectional processing to enable strategic flexibility or might instead capitalize on existing strengths by enhancing domain-specific knowledge while maintaining the successful bottom-up monitoring pattern.

### Lifelong Learning Considerations

Professional learning extends across careers spanning decades, during which both individuals and their environments change substantially. Our taxonomy suggests several considerations for lifelong learning theory:

*Scenario reconfiguration.* The same individual may occupy different scenarios depending on context. A professional functioning at Scenario 17 in their area of expertise might revert to Scenario 1 when learning a substantially new domain. Theories of lifelong learning must account for such dynamic scenario shifting.

*Age-related changes.* While chronological age correlates imperfectly with expertise, certain age-related cognitive changes may affect scenario prevalence. For instance, reduced processing speed might favor S→P control-dominant patterns over P→S monitoring-dominant patterns. However, empirical research on aging and metacognition in professional contexts remains limited.

*Technology-mediated learning.* Modern professional learning increasingly involves technology-mediated feedback [15]. Our external topology framework suggests that technology can enhance learning by providing additional shortcuts (particularly O→I through learning analytics and F→I through immediate automated feedback), but current theories inadequately address how such technology-enhanced feedback integrates with internal metacognitive processes.

## Formal Concept Analysis

As an initial step in validating the taxonomy, we employed formal concept analysis (FCA) (Wille, 1982), a mathematical method for discovering hierarchical structures in data through the identification of formal concepts—maximal sets of objects sharing common attributes. FCA has been successfully applied in educational research to reveal developmental progressions and attribute dependencies [32]. The FCA revealed a richly structured developmental landscape that challenges simplistic linear progression models while identifying clear thresholds and alternative pathways to expertise. Key findings included:

1. **Self-monitoring serves as the primary developmental gateway**, present in all scenarios beyond foundational novice (S1)
2. **Multiple pathways to expertise exist**, with bidirectional integration (S17, S24) and bottom-up specialization (S19) representing viable alternatives
3. **Developmental progression involves accumulation rather than replacement** in the mainstream pathway (S1→S6→S7→S14→S17)
4. **Parallel entry represents a critical unlocking threshold** that enables subsequent sophistication in both internal architecture and external connectivity
5. **Internal architecture and external connectivity show trade-off relationships**, with rich external engagement potentially compensating for unidirectional internal processing

The lattice structure supports targeted intervention design by:

- Identifying teachable transition points (4 major thresholds)
- Revealing prerequisite attribute chains
- Exposing alternative developmental routes for different learner profiles
- Suggesting diagnostic criteria for scenario classification

The FCA exposed five critical gaps in current metacognition theories: coordination mechanisms, exit pattern flexibility, feedback channel relationships, bidirectionality stability, and architecture-connectivity trade-offs. These gaps define a research agenda for advancing metacognition science beyond descriptive frameworks toward predictive, computationally specified models. The practical implications of the developmental pathways and transition thresholds specified in the taxonomy potentially enable diagnostic assessment to locate learners within the scenario space, targeted interventions addressing specific threshold crossings, individualized developmental planning recognizing multiple valid routes to expertise, and organizational learning design supporting diverse professional learning trajectories. The full FCA analysis is available for review upon request.

## Testable Predictions

**Research Question 1**: Does O→I (self-monitoring of outputs to shape inputs) capability serve as necessary gateway to Tier-Developing status? **Method**: Longitudinal study tracking learners' acquisition of self-monitoring and subsequent metacognitive development. **Expected Finding**: Learners who develop accurate self-monitoring will transition from Tier-Novice to Tier-Developing more rapidly than those who don't

**Research Question 2**: Can experts be reliably classified into integrated (S17), specialist (S19), and knowledge-driven (S24) subtypes? **Method**: Cross-sectional expert study with process-tracing measures distinguishing internal architecture and exit patterns **Expected Finding**: Expert subtypes will show distinct performance profiles on domain-specific versus transfer tasks

**Research Question 3**: How do learners develop capacity for simultaneous P-S activation? **Method**: Microgenetic study during Transition 3 (S7→S14) using think-aloud and eye-tracking **Expected Finding**: Successful transitioners will show evidence of chunking strategies or working memory enhancement

**Research Question 4**: Once P⇄S integration is achieved, under what conditions does it persist versus revert to unidirectional processing? **Method**: Intervention study providing P⇄S training, then examining retention under varying cognitive load and domain transfer conditions **Expected Finding**: Bidirectionality will persist in familiar domains but may revert under high load or substantial domain shifts

**Research Question 5**: What is the actual prevalence of each scenario in professional populations? **Method**: Large-scale assessment using validated instruments to classify working professionals **Expected Finding**: S7 will be modal for developing professionals; S17/S19/S24 will show roughly equal prevalence among experts

## Limitations

This theoretical analysis is subject to several important limitations that constrain interpretation and application:

***Inference from empirical patterns.*** The filtering decisions reducing 216 scenarios to 24 priorities are *inferred from* empirical patterns in research rather than *directly established by* controlled experiments systematically comparing all possible scenarios. While each elimination was grounded in published evidence, the specific filtering choices involve theoretical judgment. Alternative filtering criteria could yield somewhat different priority sets.

***Measurement constraints.*** The third filter explicitly acknowledged that current instruments cannot distinguish fine-grained temporal sequences or micro-level processing differences. This means that some scenarios eliminated as empirically indistinguishable might represent meaningfully different neuropsychological processes that currently lack adequate measurement methods. Future advances in assessment technology (such as process tracing, eye-tracking, or neurocognitive methods) might restore some eliminated scenarios to empirical viability.

***Cultural and contextual generalization.*** Our analysis assumes Western professional contexts characterized by individual autonomy, feedback-seeking norms, and explicit metacognitive reflection. Professional learning in other cultural contexts may exhibit different scenario distributions. For example, collectivist cultures that emphasize social learning might show a higher prevalence of scenarios with strong E→I links (socially mediated feedback) compared to individualistic cultures. The taxonomy's generalizability requires empirical testing across diverse contexts, findings that may accumulate over time.

***Domain specificity.*** Professional learning in technical domains (engineering, software development) may differ systematically from social domains (management, counseling) or creative domains (design, marketing). While we have attempted to identify domain-general patterns, the relative frequency of scenarios likely varies by professional domain. The five focal scenarios analyzed in depth may not equally represent all professional fields.

***Developmental timing.*** Our taxonomy identifies priority scenarios and suggests trajectories, but does not specify the time scales over which transitions occur. One professional might progress from Scenario 1 to Scenario 7 within months, while another might require years. The taxonomy provides qualitative developmental progressions but lacks quantitative temporal specification.

***Simplification of complex processes.*** The six-node model necessarily simplifies the complexity of human metacognitive processes. For example, "Structures" aggregates diverse forms of metacognitive knowledge (person, task, strategy knowledge; declarative, procedural, conditional knowledge) into a single node. "Processes" encompasses numerous cognitive operations. While this simplification enables systematic macroanalysis, it may obscure important within-node distinctions needed to understand meso and micro levels. We pointed out some of the shortcomings in the analysis (e.g. the approach did not distinguish structures as content-generators versus regulators), but there may be other simplifications that need to be discovered and clarified with computational specifications.

*Static representation of dynamic processes.* The scenario taxonomy represents metacognitive patterns at discrete time points, but actual professional learning involves continuous, dynamic processes. Real learning episodes may flow through multiple scenarios or exhibit hybrid characteristics. The taxonomy should be understood as capturing recurrent patterns within continuous processes rather than fixed states.

Despite these limitations, the systematic approach and empirical grounding provide a foundation for advancing metacognition theory beyond current frameworks' primarily descriptive accounts toward more integrative and testable models.

## Conclusion

This paper presented a refined taxonomy of metacognitive learning scenarios generated through systematic integration of major theoretical frameworks into a six-node open systems model, as implemented in the DOT Framework Coach [10]. Through constraint-based filtering informed by empirical evidence from published research, we identified 24 priority scenarios warranting theoretical and empirical attention. Formal concept analysis of five focal scenarios spanning novice to expert functioning revealed critical theoretical gaps in current frameworks, particularly regarding: (1) the coordination of multiple monitoring sources, (2) the relationship between internal metacognitive architecture and external feedback topology, (3) the mechanisms enabling transitions between developmental stages, (4) the distinction between process-driven versus knowledge-driven output generation, and (5) the existence of multiple viable trajectories to expert performance.

These findings have direct implications for advancing metacognition theory. Current frameworks require extension to address **monitoring multiplicity** (how learners integrate diverse monitoring sources), **architecture-environment interactions** (how internal processing patterns interact with external feedback structures), **developmental mechanisms** (precise processes enabling scenario transitions), **output generation modes** (distinguishing Processes-based vs. Structures-based performance), and **trajectory diversity** (acknowledging multiple pathways to expertise).

For professional development practice, the taxonomy enables targeted intervention design based on diagnostic assessment of learners' current scenarios and desired developmental trajectories. Rather than generic "metacognitive training," practitioners can design specific interventions to facilitate scenario transitions (developing P⇄S bidirectionality, adding self-monitoring shortcuts, enhancing feedback utilization).

Future research should empirically validate the taxonomy through studies examining: (1) the actual frequency distribution of scenarios in representative professional populations, (2) the effectiveness of scenario-targeted interventions, (3) longitudinal developmental trajectories showing transition patterns and time scales, (4) measurement instruments capable of reliably distinguishing priority scenarios, and (5) theoretical extensions addressing the gaps identified through deep scenario analysis. Such research would establish metacognition science on firmer empirical and theoretical foundations, advancing both basic understanding and practical application in lifelong professional learning contexts.

**Funding Declaration:** No funding was received to assist with the preparation of this manuscript.

**Competing Interest Declaration:** The authors declare that they have no conflicts of interest.

**References**

[1] J. H. Flavell, "Metacognition and cognitive monitoring: A new area of cognitive-developmental inquiry," *Amer. Psychologist*, vol. 34, no. 10, pp. 906–911, 1979. doi: 10.1037/0003-066X.34.10.906

[2] T. O. Nelson and L. Narens, "Metamemory: A theoretical framework and new findings," in *Psychology of Learning and Motivation*, vol. 26. New York, NY, USA: Elsevier, 1990, pp. 125–173. doi: 10.1016/S0079-7421(08)60053-5

[3] A. L. Brown, "Metacognition, executive control, self-regulation, and other more mysterious mechanisms," in *Metacognition, Motivation, and Understanding*, F. E. Weinert and R. H. Kluwe, Eds. Hillsdale, NJ, USA: Lawrence Erlbaum, 1987, pp. 65–116.

[4] G. Schraw and R. S. Dennison, "Assessing metacognitive awareness," *Contemp. Educ. Psychol.*, vol. 19, no. 4, pp. 460–475, 1994. doi: 10.1006/ceps.1994.1033

[5] J. Decius, "Workplace learning: A review and integration," in *The Wiley Handbook of Vocational Education and Training*, S. Billett, J. E. Tones III, and C. Harteis, Eds. Hoboken, NJ, USA: Wiley, 2024, pp. 193–211.

[6] S. I. Tannenbaum and M. A. Wolfson, "Informal (field-based) learning," *Annu. Rev. Organ. Psychol. Organ. Behav.*, vol. 9, no. 1, pp. 391–414, 2022. doi: 10.1146/annurev-orgpsych-012420-083050

[7] A. F. D. Kittel and T. Seufert, "It's all metacognitive: The relationship between informal learning and self-regulated learning in the workplace," *PLOS ONE*, vol. 18, no. 5, Art. no. e0286065, 2023. doi: 10.1371/journal.pone.0286065

[8] A. Koriat, "Metacognition and consciousness," in *The Cambridge Handbook of Consciousness*, 1st ed., P. D. Zelazo, M. Moscovitch, and E. Thompson, Eds. Cambridge, U.K.: Cambridge Univ. Press, 2007. doi: 10.1017/CBO9780511816789.012

[9] P. H. Winne and A. F. Hadwin, "Studying as self-regulated learning," in *Metacognition in Educational Theory and Practice*, D. J. Hacker, J. Dunlosky, and A. C. Graesser, Eds. Mahwah, NJ, USA: Lawrence Erlbaum, 1998, pp. 277–304.

[10] M. E. Azukas and D. Gibson, "Co-Intelligence in the Classroom: The DOT Framework for AI-Enhanced Teaching and Learning. AI Enhanced Learning, 1(2), 269-293. Association f," AIEL, vol. 1, no. 2, pp. 269–293, 2025, doi: https://doi.org/10.70725/846858waolmj.

[11] J. H. Flavell, "Speculations about the nature and development of metacognition," in *Metacognition, Motivation and Understanding*, F. E. Weinert and R. H. Kluwe, Eds. Hillsdale, NJ, USA: Lawrence Erlbaum, 1987, pp. 21–29.

[12] T. O. Nelson and L. Narens, "Why investigate metacognition?" in *Metacognition: Knowing about Knowing*, J. Metcalfe and A. P. Shimamura, Eds. Cambridge, MA, USA: MIT Press, 1994.

[13] J. Pearl and D. Mackenzie, *The Book of Why: The New Science of Cause and Effect*. New York, NY, USA: Basic Books, 2018. [Cited in manuscript; full publication details not provided in source document.]

[14] D. Gibson and D. Ifenthaler, *Computational Learning Theories: Models for Artificial Intelligence Promoting Learning Processes*. Cham, Switzerland: Springer Nature, 2024. doi: 10.1007/978-3-031-65898-3

[15] M. Rivera, L. Qiu, S. Kumar, and T. Petrucci, "Are Traditional Performance Reviews Outdated? An Empirical Analysis on Continuous, Real-Time Feedback in the Workplace," Information Systems Research, vol. 32, no. 2, pp. 517–540, Jun. 2021, doi: 10.1287/isre.2020.0979.

[16] M. Csikszentmihalyi, *Creativity: Flow and the Psychology of Discovery and Invention*. New York, NY, USA: HarperCollins, 1996.

[17] T. Kuhn, *The Structure of Scientific Revolutions*. Chicago, IL, USA: Univ. of Chicago Press, 1962.

[18] R. A. Noe and J. E. Ellingson, "Autonomous learning in the workplace," in Autonomous Learning in the Workplace. New York, NY, USA: Routledge, 2017.

[19] D.M. Ravid, D.L. Tomczak, J.C. White, & T.S. Behrend, "EPM 20/20: A review, framework, and research agenda for electronic performance monitoring," *Journal of Management*, vol. 46, no. 1, 100–126, 2020. https://doi.org/10.1177/0149206319869435

[20] T. and K. Ely, "A meta-analysis of self-regulated learning in work-related training and educational attainment: What we know and where we need to go," *Psychol. Bull.*, vol. 137, no. 3, pp. 421–442, 2011. doi: 10.1037/a0022777

[21] C. Dignath and G. Büttner, "Components of fostering self-regulated learning among students: A meta-analysis on intervention studies at primary and secondary school level," *Metacogn. Learn.*, vol. 3, no. 3, pp. 231–264, 2008. doi: 10.1007/s11409-008-9029-x

[22] V. J. Marsick and K. E. Watkins, "Demonstrating the value of an organization's learning culture: The dimensions of the learning organization questionnaire," *Adv. Develop. Hum. Resour.*, vol. 5, no. 2, pp. 132–151, 2003. doi: 10.1177/1523422303005002002

[23] P. H. Winne, "Self-regulated learning viewed from models of information processing," in *Self-Regulated Learning and Academic Achievement: Theoretical Perspectives*, 2nd ed., B. J. Zimmerman and D. H. Schunk, Eds. Mahwah, NJ, USA: Lawrence Erlbaum, 2001, pp. 153–189.

[24] G. Colvin, *Talent Is Overrated: What Really Separates World-Class Performers from Everybody Else*. New York, NY, USA: Portfolio/Penguin, 2008.

[25] E. Panadero, "A review of self-regulated learning: Six models and four directions for research," *Frontiers Psychol.*, vol. 8, Art. no. 422, 2017.

[26] S. Li, X. Huang, T. Wang, Z. Pan, and S. P. Lajoie, "Examining the Interplay between Self-regulated Learning Activities and Types of Knowledge within a Computer-simulated Environment," *Learning Analytics*, vol. 9, no. 3, pp. 152–168, Dec. 2022, doi: 10.18608/jla.2022.7571.

[27] T. Lehmann, I. Hähnlein, and D. Ifenthaler, "Cognitive, metacognitive and motivational perspectives on preflection in self-regulated online learning," *Computers in Human Behavior*, vol. 32, pp. 313–323, 2014.

[28] N. J. Cepeda, H. Pashler, E. Vul, J. T. Wixted, and D. Rohrer, "Distributed practice in verbal recall tasks: A review and quantitative synthesis," *Psychol. Bull.*, vol. 132, no. 3, pp. 354–380, 2006. [Cited in manuscript; full publication details not provided in source document.]

[29] D. Kahneman, *Thinking, Fast and Slow*. New York, NY, USA: Farrar, Straus and Giroux, 2011.

[30] T. Reimer, J. Raeskamp, J. Rieskamp, and J. Raeskamp, "Fast and frugal heuristics," in *Encyclopedia of Social Psychology*, vol. 7, 2006. doi: 10.1016/j.psychsport.2006.06.002

[31] A. S. Reber, "Implicit learning and tacit knowledge," *J. Exp. Psychol.: Gen.*, vol. 118, no. 3, pp. 219–235, 1989. doi: 10.1037/0096-3445.118.3.219

[32] U. Priss, "Formal concept analysis in information science," *Annu. Rev. Inf. Sci. Technol.*, vol. 40, no. 1, pp. 521–543, 2006. doi: 10.1002/aris.1440400120

The following references are used in Appendix 2

[33] R. Wille, "Restructuring lattice theory: An approach based on hierarchies of concepts," in *Ordered Sets*, I. Rival, Ed. Dordrecht, The Netherlands: Springer, 1982, pp. 445–470. doi: 10.1007/978-94-009-7798-3_15

[34] B. J. Zimmerman and D. H. Schunk, Eds., *Self-Regulated Learning and Academic Achievement: Theoretical Perspectives*, 2nd ed. Mahwah, NJ, USA: Erlbaum, 2009.

[35] J. Decius, L. Decius, and S. Beausaert, "Integrating multiple theoretical perspectives on informal field-based learning: The self-regulated informal learning cycle (SILC)," *Hum. Resour. Dev. Rev.*, vol. 24, no. 3, pp. 247–278, 2025. doi: 10.1177/15344843241310631

[36] J. Metcalfe and A. P. Shimamura, Eds., *Metacognition: Knowing about Knowing*. Cambridge, MA, USA: MIT Press, 1994. doi: 10.7551/mitpress/4561.001.0001

[37] B. J. Zimmerman, "Attaining self-regulation," in *Handbook of Self-Regulation*. San Diego, CA, USA: Elsevier, 2000, pp. 13–39. doi: 10.1016/B978-012109890-2/50031-7

[38] A. Ericsson and R. Pool, *Peak: Secrets from the New Science of Expertise*. Boston, MA, USA: Houghton Mifflin Harcourt, 2016.

[39] D. A. Schön, *The Reflective Practitioner*. London, U.K.: Routledge, 2017. doi: 10.4324/9781315237473

[40] M. T. H. Chi, "Laboratory methods for assessing experts' and novices' knowledge," in *The Cambridge Handbook of Expertise and Expert Performance*, 1st ed., K. A. Ericsson, N. Charness, P. J. Feltovich, and

R. R. Hoffman, Eds. Cambridge, U.K.: Cambridge Univ. Press, 2006, pp. 167–184. doi: 10.1017/CBO9780511816796.010

[41] H. L. Dreyfus and S. E. Dreyfus, "Peripheral vision: Expertise in real world contexts," *Organ. Stud.*, vol. 26, no. 5, pp. 779–792, 2005. doi: 10.1177/0170840605053102

[42] G. Schraw and D. Moshman, "Metacognitive theories," *Educ. Psychol. Rev.*, vol. 7, no. 4, pp. 351–371, 1995. doi: 10.1007/BF02212307

[43] G. Hatano and K. Inagaki, "Two courses of expertise," in *Children Development and Education in Japan*, H. Stevenson, H. Azuma, and K. Hakuta, Eds. New York, NY, USA: Freeman, 1986, pp. 262–272.

## APPENDIX 1

### Glossary of Computational Configuration Symbols

| Symbol | Description |
|---|---|
| **NODE LABELS** | |
| **I** | **Input node**: Information entering the learner's cognitive system from the environment or task context |
| **P** | **Processes node**: Active cognitive and metacognitive operations, including working memory processing, strategy execution, and online task engagement |
| **S** | **Structures node**: Stable knowledge representations, including metacognitive knowledge, beliefs, standards, goals, and schemas |
| **O** | **Output node**: Generated products, behaviors, responses, or performance resulting from internal cognitive processing |
| **E** | **Environment node**: External context, including social setting, organizational culture, task conditions, and environmental responses to learner output |
| **F** | **Feedback node**: Information about output quality, performance evaluation, or environmental response that can inform subsequent learning |
| **DIRECTIONAL OPERATORS** | |
| → | Directional influence from one node to another; represents information flow, causal influence, or functional connection (e.g., P→S means "Processes inform Structures") |
| ⇄ | Bidirectional influence between two nodes; represents reciprocal information flow with both monitoring (upward) and control (downward) active simultaneously or by taking turns over time (e.g., P⇄S means full integration) |
| **SET NOTATION** | |
| **{P,S}** | Set notation indicating multiple nodes functioning as sources or targets; represents parallel or simultaneous engagement (e.g., I→{P,S} means "Input flows to both Processes and Structures") |
| **{P,S}→O** | Multiple nodes serving as joint sources for a single target; indicates output can originate from either or both Processes and Structures |
| **I→{P,S}** | Single source distributing to multiple targets; indicates input simultaneously activates both Processes and Structures |

**COMBINATION OPERATORS**

| | |
|---|---|
| + | Conjunction operator indicating multiple connections active simultaneously (e.g., "O→F→E→I + O→I" means both the full pathway and the self-monitoring shortcut are active) |
| ⊗ | Broken link operator indicating failure or absence of expected connection; represents disrupted information flow (e.g., "O⊗F" means feedback is not being received) |

**CONFIGURATION PATTERNS**

| | |
|---|---|
| **P→S** | Bottom-up pattern: Processes inform Structures through monitoring; experience-based learning where processing dynamics update metacognitive knowledge |
| **S→P** | Top-down pattern: Structures control Processes through regulation; knowledge-driven performance where beliefs and standards guide execution |
| **P⇄S** | Integrated pattern: Bidirectional monitoring and control; full self-regulation with both experience-based updating and knowledge-based guidance |

**TOPOLOGY NOTATION**

| | |
|---|---|
| **Topology 1** | Baseline external configuration: O→F→E→I only (no shortcuts); full environmental mediation of feedback |
| **Topology 2** | O→F→E→I + O→E (direct environmental influence shortcut) |
| **Topology 3** | O→F→E→I + O→I (self-monitoring shortcut) |
| **Topology 4** | O→F→E→I + F→I (direct feedback incorporation shortcut) |
| **Topology 5** | O→F→E→I + O→E + O→I (two shortcuts: environmental and self-monitoring) |
| **Topology 6** | O→F→E→I + O→E + F→I (two shortcuts: environmental and feedback) |
| **Topology 7** | O→F→E→I + O→I + F→I (two shortcuts: self-monitoring and feedback) |
| **Topology 8** | O→F→E→I + O→E + O→I + F→I (maximum connectivity; all three shortcuts active simultaneously) |

**COMPOUND EXPRESSIONS**

| | |
|---|---|
| **I→P, P→O** | Comma separates distinct connections; this example indicates Input connects to Processes AND Processes connects to Output |
| **I→P, P→S, {P,S}→O** | Complete cross-cluster specification showing entry pattern (I→P), internal arrangement (P→S), and exit pattern ({P,S}→O) |
| **SCENARIO NOTATION** | |
| **Scenario X (I→{P,S}, P⇄S, {P,S}→O, Topology #)** | Complete scenario specification format: entry pattern, internal arrangement, exit pattern, external topology |
| **CONCEPTUAL GROUPINGS** | |
| **Internal cluster** | The subsystem consisting of Processes and Structures nodes and their interconnections; represents the learner's cognitive-metacognitive architecture |
| **External cluster** | The subsystem consisting of Output, Environment, Feedback, and Input nodes and their interconnections; represents the environmental learning loop |
| **Cross-cluster connections** | The interface links between internal and external clusters: entry points (I→P and/or I→S) and exit points (P→O and/or S→O) |
| **Shortcut** | Any direct connection that bypasses intermediate nodes in the external feedback loop (O→E, O→I, or F→I) |
| **Backbone** | The mandatory external pathway O→F→E→I that all topologies include; represents the fundamental environmental feedback cycle |

### Usage Examples

**I→P, P→S, P→O, O→F→E→I + F→I**: Experiential entry (I→P), bottom-up monitoring (P→S),process-driven output (P→O), with baseline feedback plus direct feedback incorporation shortcut (F→I)

**I→{P,S}, P⇄S, {P,S}→O, Topology 8**: Parallel entry to both nodes (I→{P,S}), fully integrated internal system (P⇄S), dual-source output ({P,S}→O), with maximum external connectivity (all shortcuts active)

**I→S, S→P, S→O, O→F→E→I**: Knowledge activation entry (I→S), knowledge-driven pattern with top-down control (S→P), structure-based output (S→O), and baseline environmental feedback with no shortcuts

**Appendix 2: Taxonomy of 24 Scenarios**

| Scenario | Scenario Notation | Brief Description | Literature References |
|---|---|---|---|
| **TIER 1: NOVICE PROFESSIONAL LEARNERS** | | | |
| **1**<br>Focal analysis | I→ [P, P→S, P→] O,<br>O→F→E→I + F→I | **Foundational experience-based learning.** Entry-level pattern with experiential processing that updates metacognitive knowledge through bottom-up monitoring; direct feedback incorporation facilitates learning but no self-monitoring yet developed. | Koriat, A. (2007). *Metacognition and consciousness.* In P. D. Zelazo, M. Moscovitch, & E. Thompson (Eds.), The Cambridge Handbook of Consciousness (1st ed.). Cambridge University Press. https://doi.org/10.1017/CBO9780511816789.012 |
| **2** | I→ [P, P→S,{P,S}→] O,<br>O→F→E→I + F→I | **Integration emergence.** Novice beginning to incorporate knowledge structures in output generation alongside procedural execution; still relies on bottom-up monitoring and external feedback. | Kittel, A. F. D., & Seufert, T. (2023). *It's all metacognitive: The relationship between informal learning and self-regulated learning in the workplace.* PLOS ONE, 18(5), e0286065. https://doi.org/10.1371/journal.pone.0286065 |
| **3** | I→ [S, S→P,P→] O,<br>O→F→E→I + O→I | **Knowledge application with self-monitoring.** Top-down control pattern where existing knowledge guides processing; emergence of self-monitoring capability (O→I) marks important metacognitive milestone. | Schraw, G., & Dennison, R. S. (1994). A*ssessing Metacognitive Awareness.* Contemporary Educational Psychology, 19(4), 460–475. https://doi.org/10.1006/ceps.1994.1033 |

| **4** | I→ [P→S, {P,S},P→] O, O→F→E→I + F→I | **Parallel processing initiation.** Beginning to activate both processes and knowledge structures simultaneously upon task encounter; output still primarily process-driven. | Winne, P., & Hadwin, A. (1998). *Studying as self-regulated learning.* In D. Hacker, J. Dunlosky, & A. Graessr (Eds.), Metacognition in educational theory and practice (pp. 277–304). Lawrence Erlbaum. |
|---|---|---|---|
| **5** | I→[S, S→P, S→] O, O→F→E→I + E→I + F→I | **Rapid declarative response.** Pure knowledge-driven pattern with multiple feedback channels; characteristic of well-learned factual or procedural knowledge retrieval in professional contexts. | Brown, A. L. (1987). *Metacognition, Executive Control, Self-Regulation, and Other More Mysterious Mechanisms*. In F. E. Weinert, & R. Kluwe (Eds.), Metacognition, Motivation, and Understanding (pp. 65-116). Hillsdale: L. Erlbaum Associates. |
| **6** | I→ [P, P⇄S, P→] O, O→F→E→I + O→I | **Bidirectional emergence.** Critical developmental transition where first bidirectional integration appears alongside self-monitoring; marks shift from novice to developing professional metacognition. | Nelson, T. O. (1990). *Metamemory: A Theoretical Framework and New Findings.* In Psychology of Learning and Motivation (Vol. 26, pp. 125–173). Elsevier. https://doi.org/10.1016/S0079-7421(08)60053-5 |
| **TIER 2: DEVELOPING PROFESSIONALS** | | | |
| **7**<br>Focal Analysis | I→ [P, P⇄S, {P,S}→] O, O→F→E→I + O→I | **Core integrated learning** (Most common pattern). Fully bidirectional internal system with dual-source output and self-monitoring; represents standard developmental pattern for effective professional learning; "reflect-while-doing" configuration. | Sitzmann, T., & Ely, K. (2011). *A meta-analysis of self-regulated learning in work-related training and educational attainment: What we know and where we need to go.* Psychological Bulletin, 137(3), 421–442. https://doi.org/10.1037/a0022777 |
| **8** | I→ [{P,S}, P⇄S, P→] O, O→F→E→I + F→I + O→I | **Enhanced monitoring.** Parallel entry pattern with bidirectional processing and multiple feedback channels (both self-monitoring and direct feedback); sophisticated monitoring development. | Kittel, A. F. D., & Seufert, T. (2023). *It's all metacognitive: The relationship between informal learning and self-regulated learning in the workplace.* PLOS ONE, 18(5), e0286065. https://doi.org/10.1371/journal.pone.0286065 |
| **9** | I→[ S, S→P, {P,S}→] O, O→F→E→I + O→I | **Goal-driven with monitoring.** Knowledge-activated top-down control with dual output sources and self-monitoring; characteristic of strategic professional task performance. | Zimmerman, B. J., & Schunk, D. H. (Eds.). (2009). *Self-regulated learning and academic achievement: Theoretical perspectives* (2. ed., reprint). Mahwah, NJ, USA:Erlbaum. |
| **10** | I→ [{P,S}, P→S, {P,S}→] O, O→F→E→I + O→I + F→I | **Rich information flow.** Parallel entry, dual exit, bottom-up monitoring, and multiple feedback channels; high information exchange supporting continuous learning. | Tannenbaum, S. I., & Wolfson, M. A. (2022). *Informal (Field-Based) Learning.* Annual Review of Organizational Psychology and Organizational Behavior, 9(1), 391–414. https://doi.org/10.1146/annurev-orgpsych-012420-083050 |
| **11** | I→ [S, P⇄S, P→] O, O→F→E→I + O→I | **Knowledge activation pathway.** Bidirectional integration initiated by prior knowledge rather than direct processing; characteristic of experienced professionals encountering familiar problem types. | Flavell, J. H. (1987). Speculations about the Nature and Development of Metacognition. In F. E. Weinert, & R. Kluwe (Eds.), Metacognition, Motivation, and Understanding (pp. 21-29). Hillsdale, NJ: Lawrence Erlbaum Associates. |
| **12** | I→ [P, P⇄S, {P,S}→] O, O→F→E→I + F→I + O→I | **Multiple feedback integration.** Core integrated pattern (Scenario 7) enhanced with direct feedback incorporation; dual feedback channels support rapid calibration and adjustment. | Rivera, M., Qiu, L., Kumar, S., & Petrucci, T. (2021). *Are Traditional Performance Reviews Outdated? An Empirical Analysis on Continuous, Real-Time Feedback in the Workplace*. Information Systems Research, 32(2), 517–540. https://doi.org/10.1287/isre.2020.0979 |
| **13**<br>Focal analysis | I→ [{P,S}, S→P,P→] O, O→F→E→I + O→I + F→I | **Specialized developing pattern**. Parallel entry with top-down control and dual feedback channels; "dual-channel calibration" pattern characteristic of knowledge-intensive professional work requiring rapid self-correction. | Decius, J., Decius, L., & Beausaert, S. (2025). *Integrating Multiple Theoretical Perspectives on Informal Field-based Learning: The Self-regulated Informal Learning Cycle (SILC).* Human Resource Development Review, 24(3), 247–278. https://doi.org/10.1177/15344843241310631 |
| **14** | I→ [{P,S}, P⇄S, {P,S}→] O, O→F→E→I + O→I | **Full internal integration.** Maximum internal connectivity (bidirectional processing, parallel entry, dual exit) with baseline external feedback plus self-monitoring; upper boundary of developing professional tier. | Metcalfe, J., & Shimamura, A. P. (Eds.). (1994). *Metacognition: Knowing about Knowing*. The MIT Press. https://doi.org/10.7551/mitpress/4561.001.0001 |
| **15** | I→ [{P,S}, P→S, {P,S}→] O, O→F→E→I + F→I + O→I + O→E | **Maximum connectivity developing.** Bottom-up monitoring with parallel entry/exit and four-shortcut external topology; transitional pattern toward expert functioning with very high information exchange. | Marsick, V. J., & Watkins, K. E. (2003). *Demonstrating the Value of an Organization's Learning Culture: The Dimensions of the Learning Organization Questionnaire*. Advances in Developing Human Resources, 5(2), 132–151. https://doi.org/10.1177/1523422303005002002 |
| **16** | I→ [S, S→P, {P,S}→] O, O→F→E→I + F→I + O→I + O→E | **Top-down with full feedback.** Knowledge-driven control with all major feedback pathways active; characteristic of strategic planning and decision-making in professional contexts. | Zimmerman, B. J. (2000). *Attaining Self-Regulation.* In Handbook of Self-Regulation (pp. 13–39). Elsevier. https://doi.org/10.1016/B978-012109890-2/50031-7 |
| **TIER 3: EXPERT/ADAPTIVE PROFESSIONALS** | | | |
| **17** | I→ [{P,S}, P⇄S, {P,S}→] O, O→F→E→I + O→I + F→I | **Expert standard pattern**. Prototypical expert configuration with full internal integration, parallel entry/exit, and dual feedback channels (self-monitoring and direct feedback); top performers systematically self-observe. | Dignath, C., & Büttner, G. (2008). *Components of fostering self-regulated learning among students. A meta-analysis on intervention studies at primary and secondary school level*. Metacognition and Learning, 3(3), 231–264. https://doi.org/10.1007/s11409-008-9029-x |
| **18** | I→ [{P,S}, P⇄S, {P,S}→] O, O→F→E→I + O→E + O→I + F→I (Topology 8) | **Maximum integration.** Fully connected system with all possible internal connections and all external shortcuts active simultaneously; represents peak adaptive expertise with parallel processing of multiple feedback streams. | Ericsson, A., & Pool, R. (2016). *Peak: Secrets from the new science of expertise.* Houghton Mifflin Harcourt. |
| **19** Focal analysis | I→ [{P,S}, P→S, {P,S}→] O, O→F→E→I + O→I + F→I + O→E | **Bottom-up expert pattern**. Unidirectional bottom-up monitoring with maximum external connectivity; "specialist expert" trajectory emphasizing experience-based knowledge updating and environmental restructuring (O→E); characteristic of creative/innovative professionals. | Reber, A. S. (1989). *Implicit learning and tacit knowledge.* Journal of Experimental Psychology: General, 118(3), 219–235. https://doi.org/10.1037/0096-3445.118.3.219 |
| **20** | I→ [{P,S}, S→P, {P,S}→] O, O→F→E→I + O→I + F→I | **Top-down expert.** Knowledge-driven control with full internal engagement and triple feedback channels; characteristic of strategic consultants and senior professionals whose expertise manifests as principled decision-making. | Schön, D. A. (2017). *The Reflective Practitioner* (0 ed.). Routledge. https://doi.org/10.4324/9781315237473 |

| **21** | I→ [S, P⇄S, {P,S}→] O, O→F→E→I + O→I + F→I | **Knowledge-activated expert.** Prior knowledge triggers fully integrated processing with dual outputs and multiple feedback channels; expert pattern initiated by schema activation rather than experiential processing. | Chi, M. T. H. (2006). *Laboratory Methods for Assessing Experts' and Novices' Knowledge*. In K. A. Ericsson, N. Charness, P. J. Feltovich, & R. R. Hoffman (Eds.), The Cambridge Handbook of Expertise and Expert Performance (1st ed., pp. 167–184). Cambridge University Press. https://doi.org/10.1017/CBO9780511816796.010 |
|---|---|---|---|
| **22** | I→ [P, P⇄S, {P,S}→] O, O→F→E→I + O→E + O→I + F→I (Topology 8) | **Process-initiated expert.** Experience-based entry point with maximum integration; characteristic of expert practitioners whose expertise manifests in real-time adaptive performance during complex task execution. | Dreyfus, H. L., & Dreyfus, S. E. (2005). *Peripheral Vision: Expertise in Real World Contexts*. Organization Studies, 26(5), 779–792. https://doi.org/10.1177/0170840605053102 |
| **23** | I→ [{P,S}, P⇄S, P→] O, O→F→E→I + O→I + F→I + O→E | **Monitoring-focused expert.** Full integration with parallel entry but process-driven output; exceptional metacognitive awareness with environmental influence; may characterize expert coaches or reflective practitioners. | Schraw, G., & Moshman, D. (1995). *Metacognitive theories*. Educational Psychology Review, 7(4), 351–371. https://doi.org/10.1007/BF02212307 |

| **24** Focal analysis | I→ [{P,S}, P⇄S, S→] O, O→F→E→I + O→I + F→I + O→E | **Integrated expert, knowledge-driven output** Maximum metacognitive sophistication with knowledge-based output generation; "adaptive expert" pattern characteristic of thought leaders, strategic planners, and professionals whose primary outputs are conceptual frameworks rather than procedural executions. | Hatano, G., & Inagaki, K. (1986). *Two courses of expertise*. In H. Stevenson, H. Azuma, & K. Hakuta (Eds.), Children development and education in Japan (pp. 262-272). New York: Freeman. |
|---|---|---|---|

**Notes:**

1. Topology 8 refers to the maximally connected external configuration: O→F→E→I + O→E + O→I + F→I (all three shortcuts active).
2. Focal analysis indicates scenarios subjected to deep theoretical analysis in the main text (Scenarios 1, 7, 13, 19, and 24).
3. Literature references represent foundational or supporting sources for the theoretical patterns described. Many scenarios represent novel integrations not explicitly discussed in prior literature, though their constituent elements are well-established.
4. The progression from Tier 1 through Tier 3 reflects developmental trajectories, though alternative pathways exist (see main text Discussion section).